\documentstyle[11pt,aaspp4]{article}
 
\slugcomment{{\it Astrophysical Journal}, {\bf 540}, Sep 10, 2000, in press}
\lefthead{Elmegreen and Hunter}
\righthead{A Pressure Anomaly in Irregular Galaxies} 
\def\xx{\enspace\enspace}
\def\et{et al.}

\begin{document}
 
\title{A Pressure Anomaly for HII Regions
in Irregular Galaxies}

\author{Bruce G.\ Elmegreen}
\affil{IBM T.\ J.\ Watson Research Center, PO Box 218, Yorktown Heights,
New York 10598 USA; \\bge@watson.ibm.com}
 
\author{Deidre A.\ Hunter}
\affil{Lowell Observatory, 1400 West Mars Hill Road, Flagstaff, Arizona 86001
USA; \\dah@lowell.edu}

\begin{abstract}
 
The pressures of giant HII regions in 6 dwarf Irregular galaxies are
found to be a factor of $\sim10$ larger than the average pressures of
the corresponding galaxy disks, obtained from the stellar and gaseous
column densities. This is unlike the situation for spiral galaxies where
these two pressures are approximately equal. Either the HII regions in
these dwarfs are all so young that they are still expanding or there is
an unexpected source of disk self-gravity that increases the background
pressure.

We consider first whether any additional self-gravity might come from
disk dark matter that is either cold H$_2$ gas in diffuse or
self-gravitating clouds with weak CO emission, or is the same material
as the halo dark matter inferred from rotation curves. The H$_2$
solution is possible because cold molecular clouds would be virtually
invisible in existing surveys if they were also CO-weak from the low
metal abundances in these galaxies. Cosmological dark matter might be
possible too because of the relatively large volume fraction occupied by
the disk within the overall galaxy potential. There is a problem with
both of these solutions, however: the vertical scale heights inferred
for Irregular galaxies are consistent with the luminous matter alone.
The amount of disk dark matter that is required to explain the high HII
region pressures would give gas and stellar scale heights that are 
too small. 

The anomalous pressures in star-forming regions are more likely the
result of local peaks in the gravitational field that come from large
gas concentrations. These peaks also explain the anomalously low average
column density thresholds for star formation that were found earlier
for Irregular galaxies, and they permit the existence of a cool HI
phase as the first step toward dense molecular cores.  The evidence
for concentrations of HI in regions of star formation is summarized;
the peak column densities are shown to be consistent with local pressure
equilibrium for the HII regions.  Strongly self-gravitating star-forming
regions should also limit the dispersal of metals into the intergalactic
medium.

The third possibility is that all of the visible HII regions in these
dwarf galaxies are strongly over-pressured and still expanding.  The mean
time to pressure equilibrium is $\sim15$ times their current age, which
implies that the observed population is only 7\% of the total if they live
that long; the rest are presumably too faint to see. The expansion model
also implies that the volume filling factor can reach $\sim100$ times the
current factor, in which case faint and aging HII regions should merge
and occupy nearly the entire dwarf galaxy volume.  This would explain the
origin of the giant HI shells seen in these galaxies as the result of old,
expanded HII regions that were formerly driven by OB associations. The
exciting clusters would now be so old and dispersed that they would not be
recognized easily.  The shells are still round because of a lack of shear.

\end{abstract}

\keywords{galaxies: irregular --- stars: formation --- HII Regions}

\indent received 16 February 2000; accepted  13 April 2000
 
\section{Introduction}

Dwarf Irregular galaxies often have relatively large and bright HII
regions indicative of active star formation in dense neighboring clouds.
Normal spiral galaxies have equally bright HII regions, but the background
stellar disks in spiral galaxies are usually bright too, giving the
HII regions less contrast than they have in dwarfs. This difference
leads to the well-known patchy and apparent starburst quality of dwarf
galaxies. Here we point out that the difference also translates into a
pressure anomaly that may have important implications for the distribution
of matter or the origin of giant shells.

The pressures in the largest HII regions of 6 dwarf Irregular galaxies
are found to be $\sim10$ times larger than the pressures from the average
gravitational binding of the disk in the vertical direction, unlike
the case for spiral galaxies where these two pressures are comparable
(Sects. \ref{sect:press}, \ref{sect:p}).  Either the HII regions are
actively expanding to fill the disk volume and part of the halo (their
volumes have to increase by a factor of $\sim100$ before they reach
pressure equilibrium), or there is additional mass in the star-forming
regions that increases the ambient disk pressure.  We show below that
this additional mass corresponds to a factor of $\sim3$ enhancement in
column density if it is in the form of gas, and a factor of $\sim10$
enhancement in column density if it is in the form of non-interacting
cosmological particles.

We first consider the possibility that there might be additional mass in
a pervasive dark form (Sects. \ref{sect:ddm}).  The factor of $\sim10$
needed to account for the disk pressure anomaly is comparable to the
mass factor needed to account for dwarf galaxy rotation curves. Disk
dark matter was previously considered for dwarf galaxies based on the
gravitational instability threshold for star formation and on the need
for a cool atomic phase of interstellar matter (Hunter, Elmegreen, \&
Baker 1998; hereafter Paper I). There seems to be a problem with scale
heights in this case, however: if pervasive dark matter in any form
explains the HII region pressure anomaly, then the interstellar scale
heights would be 3 to 10 times smaller than the values usually inferred
for dwarf galaxies (Sect. \ref{sect:prob}).

A better solution is to have the visible matter distributed
non-uniformly, with large concentrations of gas around the HII regions
and relatively small column densities elsewhere (Sect. \ref{sect:local}).
Observations of dwarfs already suggest that most of the inner-disk HI
gas is concentrated directly in the regions of star formation (see also
van Zee et al.  1997). This raises the pressure locally, and provides
some resistance to HII region expansion. A comparison with the gas
distributions in spiral galaxies (Sect. \ref{sect:spiral}) suggests that
the same physical processes may be at work in the formation of clouds,
but the presence of shear and spiral waves in the larger systems makes the
morphology, and possibly the ages, of the equivalent gas concentrations
different.

Any systematic concentration of gas in the immediate vicinity of star
formation raises the possibility that enriched outflows from dwarf
galaxies may be less likely than previously believed (Dekel \& Silk
1986). The outflow would not feel the weak potential of the galaxy
at first, allowing its easy escape, but the deep potential well of
the star-forming concentration and the enhanced drag from all of the
additional gas around it. Only if the outflow can get around this local
resistance may it get a chance to leave the galaxy.

Another possibility is that the HII regions are only the brightest and
highest-pressure members of a larger population of HII regions that are
still expanding into an anomalously low pressure interstellar medium
(Sect. \ref{sect:shell}).  This is consistent with our detection
limitations, and also has interesting implications for the origin of
giant HI shells.

Distinctions between these possibilities can be determined by independent
measures of the turbulent pressures and disk dark matter contents of dwarf
galaxies, by observations of the kinematics and morphology of extremely
faint HII regions, and by any correlation between old and dispersed OB
associations and giant HI shells.

\section{Pressures of HII regions}
\label{sect:press}

The pressures of the largest HII regions in a galaxy, if they are
not too faint to be seen, should be a good indicator of the ambient
interstellar pressure. HII regions expand quickly when they are young,
but they decelerate after a few internal sound crossing times and expand
more slowly after this.  In a galactic population with a wide age spread,
most of the HII regions are close to their largest sizes and lowest
pressures (Sect. 3). Many of them should be nearly stalled in pressure
equilibrium, although a few of the largest might still break out of the
disk when their radii exceed a scale height.

A primary concern is whether the pressure-equilibrium HII regions are
too faint to be observed.  If the ambient pressure is high, then a large
fraction of the HII regions in a galaxy should be observable because their
limiting sizes are relatively small and their final emission measures
relatively high.  If the ambient pressure is low, then the expansion to
pressure equilibrium takes a very long time and most of the HII regions
in a galaxy may be too faint to see. Such a hidden population of old HII
regions would have interesting consequences for interstellar structure.

\begin{figure}
\vspace{7.0in}
\includegraphics{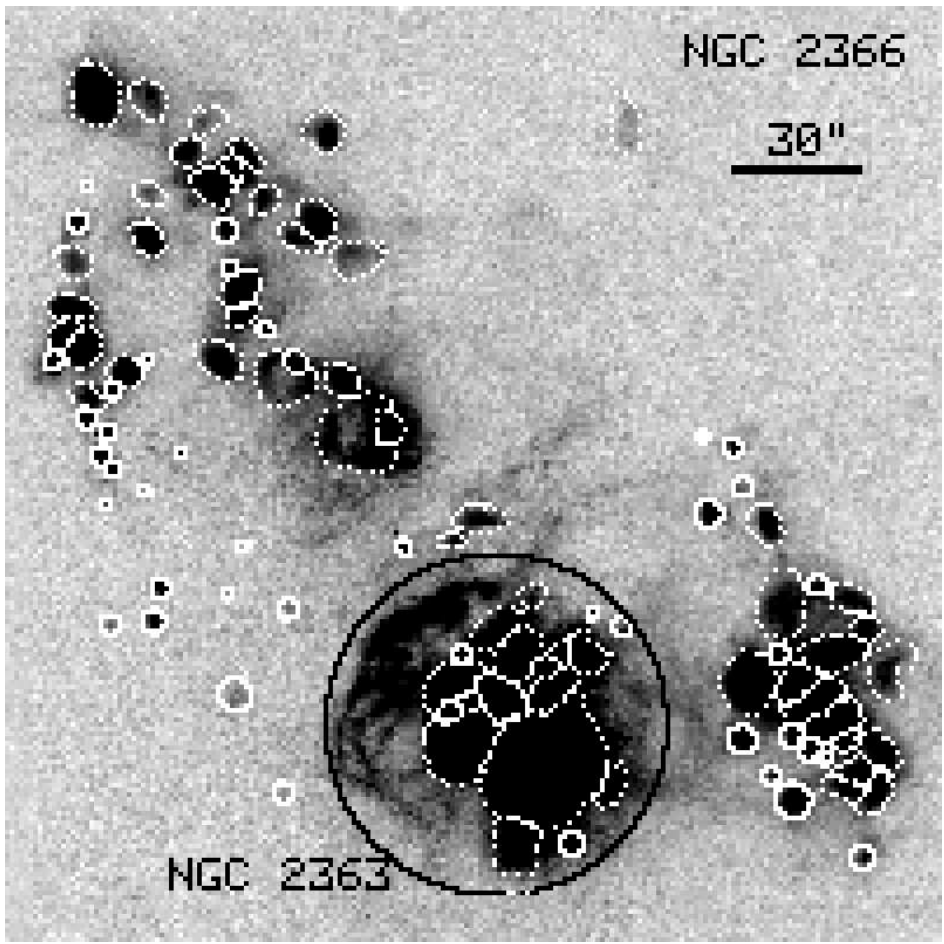}
\caption{The central portion of NGC 2366 in H$\alpha$ with HII regions and 
the giant complex NGC 2363 outlined. The image was taken with the 1.8 m
Perkins telescope at Lowell Observatory 1994 March with a TI CCD
and a 30 \AA\ FWHM filter.
Stellar continuum has been subtracted to leave nebular emission.}
\label{fig:2366}
\end{figure}

Figure \ref{fig:2366} shows an H$\alpha$ image of NGC 2366. Prominent
HII regions as defined by Youngblood \& Hunter (1999) are outlined,
and a circle is drawn around the supergiant HII complex NGC 2363 (which
was recently studied by Drissen et al. 1999). Many of the HII regions
have well-defined edges, making them {\it appear to be bounded by some
external pressure}. The arcs or partial shells of emission around the
complex have the same character.

\begin{figure}
\vspace{7.0in}
\includegraphics{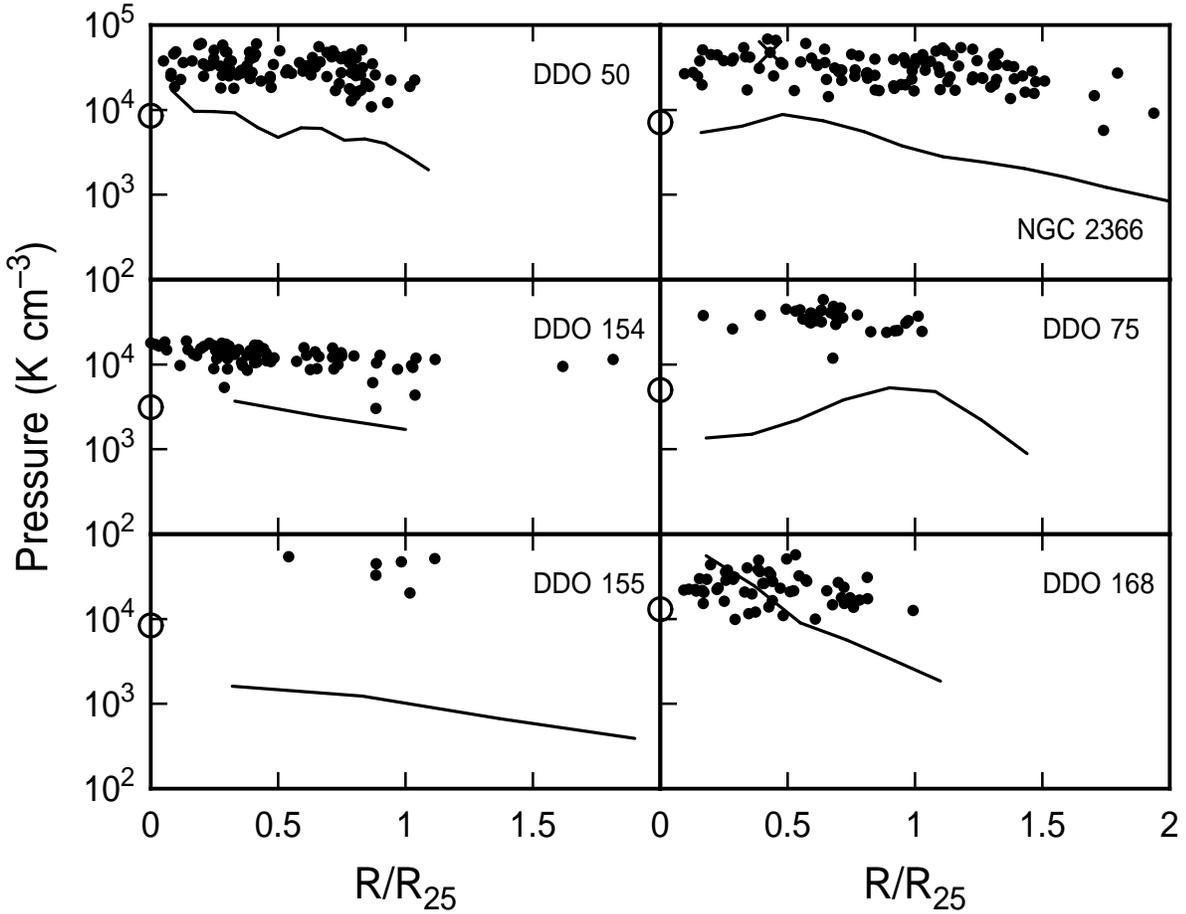}
\caption{Pressures of HII regions and the ambient disks for
dwarf Irregular galaxies.  
The points are individual HII regions in each galaxy and the cross
for NGC 2366 is the complex outlined in Fig. 1. 
The circles on the y axes are representative
detection limits. The solid
line is the average midplane pressure estimated from the HI gas
column density and the V-band surface brightness.
R is the distance from the center of the galaxy; R$_{25}$ is
the radius of the galaxy to the B-band surface brightness level
of 25 magnitudes arcsec$^{-2}$, taken from de Vaucouleurs et al.
(1991).}
\label{fig:p}
\end{figure}

We examined the radii, $R$, and H$\alpha$ luminosities, $L$, of these
HII regions in NGC 2366, and of similar HII regions in five other dwarf
Irregular galaxies, listed in Table 1. The luminosities were measured
from CCD photometry by Youngblood \& Hunter (1999) and converted to Lyman
Continuum photon counts per second by multiplying by $7\times10^{11}$
erg$^{-1}$. The result was converted to $n^2R^3$ by dividing by the
recombination rate $\alpha=2.6\times10^{-13}$ cm$^{-3}$ s$^{-1}$
(case B recombination at $10^4$ K; Osterbrock 1974). This quantity
was then divided by $4\pi R^3/3$ for the observed HII region radius,
and the square root of this gives the rms density. To get a measure
of the {\it rms thermal pressure}, we multiply the rms density by
$2.1\times10^4$, which assumes contributions from electrons, protons,
and a 10\% abundance of He at $10^4$ K. The results are plotted as points
in Figure \ref{fig:p}. The cross in the panel for NGC 2366 is the value
for the whole complex of HII regions outlined in Figure \ref{fig:2366}
(NGC 2363).

Detection limits were estimated for the HII regions in our galaxies
by Youngblood \& Hunter (1999).  They were determined for a moderately
crowded field and are only meant to be representative of limiting surface
brightness. These limits are shown in Figure \ref{fig:p} as circles on
the y-axes. They are usually at the lower borders of the HII regions
plotted, indicating that the observation have found emission down to
the detection limit. This implies that fainter HII regions may be present.

The rms thermal pressures in Figure \ref{fig:p} are not the peak
pressures in the HII regions because the densities vary as a result
of clumpy structure and because turbulent motions increase the overall
velocity dispersion in the ionized gas. The pressure can also increase
by a factor of 100 or more in the ionization front at the edge of an
embedded neutral cloud. For example, in NGC 2366, our [OII] spectra give
densities of $\sim200$ cm$^{-3}$ and 50 cm$^{-3}$ (Hunter \& Hoffman
1999) in HII regions where the rms densities calculated above equal 3.3
and 1.4, respectively. This [OII] emission is coming from an ionization
front that has a pressure much larger than the average. The rms thermal
pressures are useful primarily as a measure of the {\it minimum pressure
near the diffuse edge of the HII region}.  It is expected to be smaller
than the total pressure there by perhaps a factor of $\sim2$ because of
the exclusion of turbulent motions in our calculations.  This boundary
is an important part of the HII region for the present discussion because
it is where the expanding gas interacts with the ambient ISM.

The HII region density used for this calculation is the rms density
obtained from the emission, not the average density. The average
density in a region is always less than the rms density, so we have to
demonstrate that the pressure obtained from the rms density is a useful
concept. This is most easily done when the density fluctuations result
from supersonic turbulence, as is likely to be the case far from neutral
cloud edges. The pressure we really want is not the thermal pressure
obtained above using the rms density, which may be written $n_{rms}\mu
c_{HII}^2$ for mean atomic mass $\mu$, but the total of the thermal
and turbulent pressures, which is obtained from the average density
and the quadratic sum of the sound speed and the turbulent velocity,
$n_{ave}\mu\left(c_{HII}^2+v_{turb}^2\right)$. We can get this total
pressure using an idealized model in which a fraction $f$ of the
volume is in turbulence-compressed regions of density $n_{comp}$, and a
fraction $(1-f)$ of the volume is an interclump gas, of density $n_0$.
For non-magnetic turbulence compression, these two densities are related
by the square of the Mach number, $n_{comp}/n_0=v_{turb}^2/c_{HII}^2={\cal
M}^2$. The rms density is related also, by the equation
$n_{rms}=\left(fn_{comp}^2+[1-f]n_0^2\right)^{1/2}$. The average density
is $n_{ave}=fn_{comp}+(1-f)n_0$. From these expressions, we can get
the ratio of the desired total pressure to the rms thermal pressure
calculated above. This ratio is: \begin{equation}
{{P_{total}}\over{P_{rms,thermal}}}=
{{n_{ave}\left(c_{HII}^2+v_{turb}^2\right)}\over{n_{rms}c_{HII}^2}}=
\left(M^2+1\right){{fM^2+1-f}\over{\left(fM^4+1-f\right)^{1/2}}}.
\end{equation} Sample values of this ratio are shown in Table 2. For
$M=1$, the ratio equals 2 for all $f$, so the value is not tabulated.
The ratio exceeds 2 for $M>1$, and typically exceeds 10 for $M>5$.
Thus, the rms thermal pressures in Figure 1 are
reasonable lower limits to the total pressures in turbulent
HII regions. The
actual pressure could be larger by at least a factor of 2, depending on
the strength of the turbulent motions. 

\begin{figure}
\vspace{7.0in}
\includegraphics{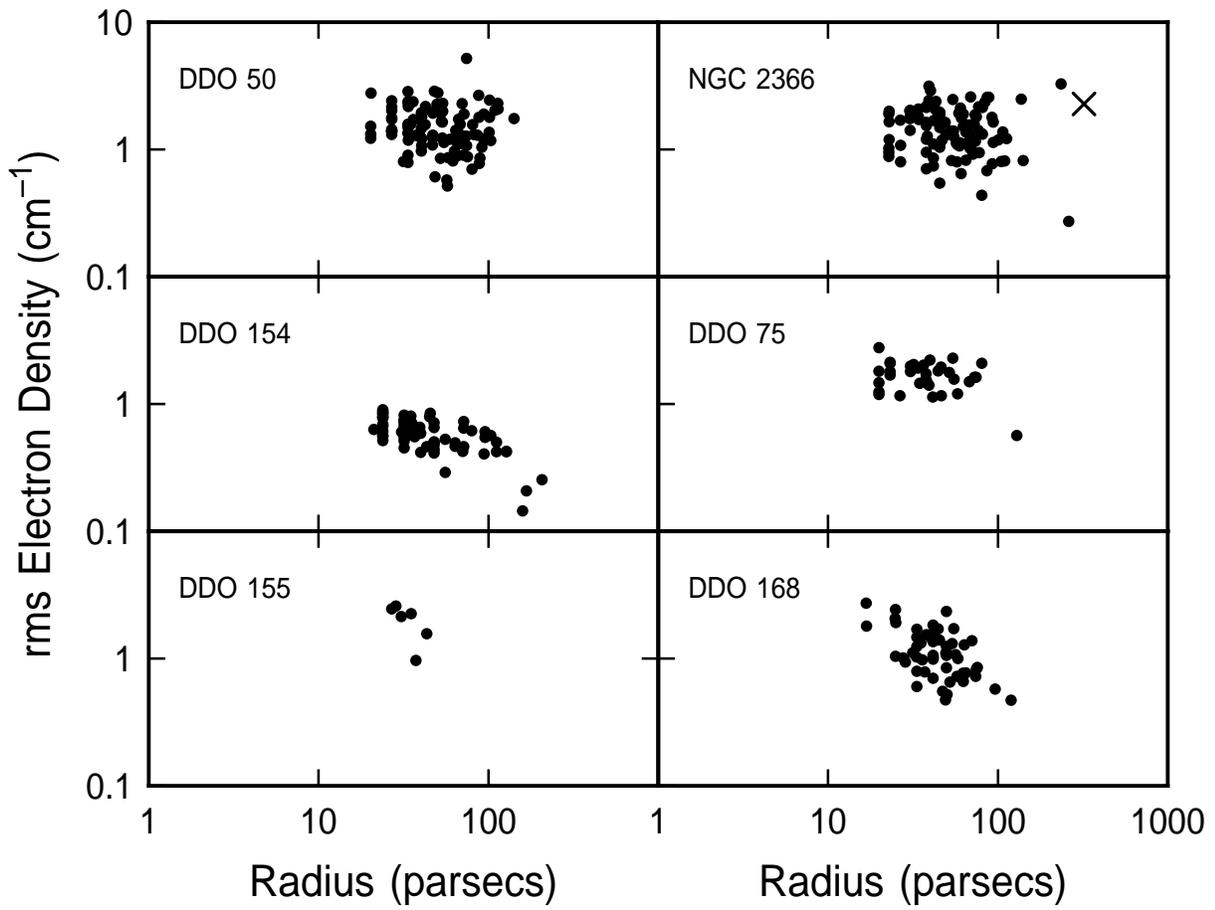}
\caption{HII region rms densities versus sizes. Only the
largest HII regions are considered here because of resolution
limitations with the observations.  These densities
and sizes are comparable to those in giant spiral galaxies.}
\label{fig:n}
\end{figure}

Figure \ref{fig:n} shows the rms densities in the HII regions versus their
radii for comparison with similar plots of Galactic HII regions made by
Habing \& Israel (1979) and of the largest HII regions in nearby galaxies,
made by Kennicutt (1984). Our survey for dwarf Irregulars includes only
the resolved HII regions, and for them, the densities are about the same
as for the HII regions with similar sizes in our Galaxy and other giant
spirals. Smaller HII regions in our sample of dwarf galaxies should have
larger densities like the smaller HII regions in giant galaxies.

The decrease in rms density with increasing HII region size is the
result of a combination of expansion for some HII regions, which
gives this correlation even for a uniform ambient medium, plus pressure
equilibrium for other HII regions, which gives this correlation when the
surrounding pressure varies from place to place.  In the latter case,
the lower pressure regions have lower-density equilibrium HII regions,
and larger Str\"omgren radii for the same luminosities.  As a result of
these effects, the variation in rms density is between a factor of 3 and
10 for the observed range of sizes.  This is a relatively small scatter,
considering the inaccuracy of the pressure derivation, but lower-density
HII regions cannot be seen with these sizes, so the small scatter is
partly an artifact of the detection limit. The scatter from one galaxy to
another is even less: most of the observable giant HII regions in these
dwarf galaxies have an rms density close to 1 cm$^{-3}$.  This is the
same as the rms density of most observable giant HII regions in normal
spirals (Kennicutt 1988; see his Fig. 6), but these have comparable
detection limits.  This similarity to giant spirals gives the impression
that the HII regions studied here are normal in terms of their densities,
sizes, and rms thermal pressures.  Any differences between dwarf galaxy
HII regions and spiral galaxy HII regions occurs below the H$\alpha$
detection limit.

\section{Mean ISM Pressures in the Galactic Disks}
\label{sect:p}

The average pressure in the local ISM is $\sim3\times10^4$ K cm$^{-3}$
from the summation of turbulent and thermal motions, magnetic fields,
and cosmic rays (Parker 1966; Boulares \& Cox 1990; Ferrara 1993). This
is comparable to the average thermal HII region pressure in the largest
HII regions in giant spiral galaxies, considering that their electron
densities are always about 1 cm$^{-3}$ (e.g., Kennicutt 1988). This
comparison leads to the sensible conclusion that the largest HII regions
have expanded to some pressure equilibrium with their environment.
Alternatively, the HII regions could still be freshly ionizing the
large-scale ambient medium around them, but because the effective rms
velocity dispersion of the general ISM is about the same as the effective
rms dispersion of an HII region, their pressures are about the same
(Kennicutt 1988).

The origin of the average ISM pressure is energy input from a combination
of supernovae, stellar winds, and other point sources, and, in the case
of the magnetic field, from differential galactic rotation too.  However,
the disk expands in the vertical direction to adjust its pressure, so
we cannot derive the pressure simply by adding up the contributions from
midplane energy sources; we also have to consider the height of the gas
layer relative to the stars.

The easiest way to estimate the total average ISM pressure is from
the self-gravitational binding energy density of the disk in the
vertical direction. This automatically includes the energy sources if
the vertical velocity dispersion of the gas is chosen correctly. The
result is a midplane pressure equal to (Elmegreen 1989) \begin{equation}
P\sim{{\pi}\over{2}}G\sigma_g\left(\sigma_g+{{c_g}\over{c_s}}\sigma_s
\right) \label{eq:p} \end{equation} where $\sigma_g$ is the total gas
mass column density, $\sigma_s$ is the stellar mass column density,
$c_g$ is the total effective vertical gas velocity dispersion, e.g.,
as it enters into an expression for the scale height, and $c_s$ is
the effective stellar vertical velocity dispersion. The gas velocity
dispersion, $c_g$, contains the magnetic ($P_B$) and cosmic ray ($P_{CR}$)
pressures and, in the notation of Parker (1966), may be written in
terms of the turbulent pressure $P_t$ and 1-D rms gas speed $c_{rms}$
as $c_g=c_{rms}\left(1+\alpha+\beta\right)$ where $\alpha=P_B/P_t$
and $\beta=P_B/P_t$ are of order unity.  The term in the parentheses
of equation \ref{eq:p} is the total column density of gas and stars
inside the gas layer. This total column density is multiplied by the
pure-gas column density, outside the parenthesis, to get the total
midplane pressure.

To evaluate the average midplane pressure for a galaxy, we need
the average gaseous and stellar column densities and some estimate
for $c_g/c_s$. The gaseous column density is taken from the radial
distribution of the average HI column density given by published radio
interferometer observations, corrected for He and heavy elements. For
the moment, we ignore the unknown contribution to $\sigma_g$ from H$_2$.
The stellar column density is taken from the radial profile of the
average V-band surface brightness and the mass-to-light ratio given by
the optical color, either as a function of radius or for the galaxy as
a whole. The surface brightness $\mu$ is converted into surface mass
density by the equation: \begin{equation} \sigma_s=4.3\times10^8\times
10^{-0.4\left(\mu-4.84\right)}{{M}\over{L}} \;\;{\rm M}_\odot \;\;{\rm
pc}^{-2} \end{equation} In this equation, the surface brightness in
mag arcsec$^{-2}$, scaled to the Sun's absolute V magnitude of 4.84
mag, is converted into a surface luminosity in L$_\odot$ per pc$^2$
by dividing by the area of 1 square arcsec at a distance of 10 pc
($4.3\times10^8=[180\times3600/\pi/10]^2$). The result is multiplied
by the mass-to-light ratio to get the surface density in M$_\odot$
pc$^{-2}$. The ratio of velocity dispersions, $c_g/c_s$ in equation
(\ref{eq:p}), is taken equal to 0.5.

Figure \ref{fig:p} shows the average midplane pressure versus radius
for the 6 dwarf Irregular galaxies in our survey.  The average
midplane pressure calculated only from the gas column density and
the V-band surface brightness is less than the HII region pressure
by about an order of magnitude for these galaxies. Table 3 gives the
averages of the pressures in the HII regions and in the disks for
radii beyond $0.5R_{25}$, and it gives the ratios of these averages.
These ratios range from $\sim5$ to $\sim15$ in all cases but DDO155,
where it is $\sim55$. Evidently, the rms thermal pressures in all of
the HII regions exceed the average midplane pressures calculated from
disk self-binding. If we consider that most HII regions also have an
equal or greater contribution to pressure from turbulent motions, and
that the pressures at some of the ionized edges can be larger than the
average pressure if there are ionization fronts (see Table 2), then we
conclude that the observable HII regions are typically over-pressured
by more than a factor of ten. In what follows, we consider this factor
of 10 to be representative of the measured HII region over-pressures in
dwarf Irregular galaxies.

In giant spiral galaxies, the average midplane pressure calculated in the
same way is essentially the same as the rms thermal pressure of giant HII
regions, namely several $\times10^4$ K cm$^{-3}$. The HI column densities
are about the same in the two galaxy types too. The difference from the
present result is that the average surface brightness of giant spirals
is several magnitudes larger than the surface brightnesses of dwarfs,
and this factor of $\sim10$ translates directly into an ISM pressure
difference. Note that even if we took $c_g/c_s\sim1$ (Bottema, Shostak \&
van der Kruit 1986), which is the maximum likely value, the HII regions
in dwarfs would still be over-pressured.

The HII regions in these dwarf galaxies are either confined by some
additional background pressure that was not included in the average
disk pressure determined above, or they are expanding into a much
lower-pressure interstellar medium and disappearing at the detection
threshold.  We consider in the next three sections the first of these
options: whether star formation in dwarf galaxies might typically occur
in anomalously high pressure clouds compared to the rest of the disk.
Then we consider the implications of the second option, a prolonged
expansion of HII regions into low pressure galaxies.

\section{Disk Dark Matter}
\label{sect:ddm}

The ambient pressure in a galaxy disk depends entirely on the column
density of material.  If the ambient pressure is larger than the average
column density suggests, then there has to be an additional mass in
the disk that has not yet been counted. We first consider whether this
additional mass can be dark matter in some form.

The possibility that a substantial fraction of a spiral galaxy's dark
matter is in its disk has been discussed extensively over the last
several years, particularly in regard to unseen molecular gas. Lequeux,
Allen, \& Guilloteau (1993) found cold CO in the outer Galaxy disk and
suggested it might account for a significant fraction of the dark matter
inferred from the rotation curve. Anomalous extinction in the outer part
of the SMC led to the same suggestion by Lequeux (1994). Pfenniger,
Combes \& Martinet (1994) reviewed the evidence for large amounts of
unseen molecular material in other galaxies. Pfenniger \& Combes (1994)
provided a model for how it might be hidden, while Combes \& Pfenniger
(1997) gave several suggestions for how it might be observed. In M31,
cold CO was inferred from observations of the inner disk by Loinard,
Allen, \& Lequeux (1995).

Dark molecules might also be present in galactic halos. In the thick
disk of NGC 891, a substantial amount of warm H$_2$ (150-230K), enough
to account for the rotation curve, was directly observed by Valentijn \&
van der Werf (1999). Dark molecular gas in the Galactic halo was discussed
by de Paolis, et al. (1995a,b) as an explanation for microlensing events,
and by Kalberla, Shchekinov, \& Dettmar (1999) as an explanation for
the background gamma ray emission.

This picture of dark baryonic matter is not without problems, however.
Early attempts to detect substantial amounts of cold H$_2$ in our
Galaxy were unsuccessful (Evans, Rubin, \& Zuckerman, 1980; Wilson,
\& Mauersberger 1994), and the COBE flux in the Milky Way is accounted
for entirely by dust associated with known gas (Sodroski et al. 1997).
Surface photometry of other galaxies also shows too little extinction
for the dynamical dark matter to be in the form of dispersed gas (D.
Elmegreen 1980; Beckman et al. 1996; Berlind et al. 1997; Pizagno \&
Rix 1998). Moreover, the expected effects of hidden cold disk matter on
disk stability and the gaseous scale height are unobserved (Knapp 1987;
Merrifield 1992; Malhotra 1994, 1995; Elmegreen 1996; Olling \& Merrifield
2000). In the solar neighborhood, the vertical scale height of both gas
(van der Kruit \& Shostak 1984) and stars (Kuijken \& Gilmore 1989,
1991) can be explained without much disk dark matter at all. The same
is true for the edge-on galaxy NGC 891 (van der Kruit 1981). However,
Olling (1996) obtained a different result for NGC 4244, suggesting that
the dark matter has an aspect ratio of $\sim0.2$ unless the gas motions
are non-isotropic, and Becquaert \& Combes (1997) proposed a flattened
distribution for NGC 891 considering also a warp.

There is probably a substantial amount of warm H$_2$ in galaxies that is
not associated with CO emission or any other currently-observed emitting
phase, as is the case for many of the diffuse clouds in the local ISM
(Spitzer \& Jenkins 1975), and there may also be cold CO+H$_2$ gas that
is too poorly excited to emit in CO. The molecular fraction of the ISM
is high in the inner regions of spiral galaxies anyway (Honma, Sofue, \&
Arimoto 1995), presumably because of the high dust abundance to enhance
molecule formation (Smith et al.  2000), and because of the high pressure
for enhanced self-shielding (Elmegreen 1993). If it is high in dwarf
galaxies, which have low metal abundances and lower stellar column
densities, making it more difficult to have high pressures, then the
reason for this would have to be different.

The situation is indeed very different in dwarf galaxies. There the
radiation field can be so low in regions far from star formation that
cold molecular gas might be more prevalent, particularly in large,
low-density self-gravitating clouds.  A similar situation might occur in
the far-outer regions of spiral galaxies. Secondly, dwarf galaxies have
such a low mass that the virial velocity or rotation speed is only a few
times the gaseous velocity dispersion. In this situation, the two-fluid
instability (Jog \& Solomon 1984) couples the dark matter and the gas,
enhancing the effective mass of the latter (Paper I). This gives the disk
a larger self-gravity and lower Toomre Q value than it would have from
the visible gas and stars alone. It also increases the midplane pressure
(Paper I).

The amount of disk dark matter necessary to equilibrate the back
pressure of the ISM on an HII region depends on the form of this matter.
If it contributes to the pressure directly, in addition to the disk
self-gravity, then it acts like an additional term in $\sigma_g$, and only
a factor of $\sim3-4$ times the HI column density is necessary to increase
$P$ in equation (\ref{eq:p}) by the required factor of $\sim10$.  This is
because $\sigma_g$ occurs both outside and inside the parenthesis: the
one outside acts like a density in the expression for pressure (actually
it is the midplane gas density times twice the scale height), and the one
inside determines the squared velocity dispersion (in the expression for
scale height). If, on the other hand, the disk dark matter contributes
only to the self-gravity, then it acts in equation (\ref{eq:p}) like the
stellar surface density, $\sigma_s$, and a factor of $\sim10$ times the
current total disk column density is needed to give pressure equilibrium.

In either case, there is no known material that is present in these
galaxy disks {\it at a pervasive level} of $\sim3-10$ times the average
HI column density or stellar surface density. If the unseen material is
distributed gas, then the associated dust would increase the opacity of
the disk to unacceptable levels. If it is stellar, then the mass function
would require an unusually large peak at the brown dwarf mass in order to
hide 10 times the stellar luminosity from the optical bands. It has to be
in a dark and compact, unresolved state, such as tiny clouds (Pfenniger
\& Combes 1994), old degenerate stars, or elementary particles. In this
case, it would seem to be unable to contribute to the ISM back-pressure
directly, and so would need the factor of $\sim10$ enhancement rather
than the factor of $\sim3$.

One possibility is that the dark matter is distributed and gaseous but
only in the outer disk, beyond the HII regions. Then it may show up as
extinction for background galaxies (Lequeux 1994), but not contribute
to the disk pressure directly. In this case, the gravity from the outer
disk would still contribute to the perpendicular force in the inner disk,
and therefore enter into the parenthesis of equation (\ref{eq:p}), but
a factor of $\sim10$ mass enhancement would be needed, not just $\sim3$
for inner-disk gas.

A factor of $\sim10$ enhancement in dwarf galaxy disk dark matter
is close to the factor needed for dwarf galaxy dark matter overall,
as determined from the rotation curves (Kormendy 1988; Carignan \&
Freeman 1988; Carignan \& Beaulieu 1989; van Zee et al. 1996; Swatters
1999). If we consider the large disk thicknesses for these galaxies
(Hodge \& Hitchcock 1966; van den Bergh 1988), and the correspondingly
large disk aspect ratios ($0.58$ in Staveley-Smith et al. 1992; $\sim0.6$
in Carignan \& Purton 1998), then perhaps half of the total dark matter
is within the disk anyway. Thus cosmological dark matter could solve the
pressure problem: the fraction of it that happens to reside in the disk
volume would participate in the dynamics of the disk, contributing to the
overall disk self-gravity and therefore the ambient ISM pressure, and
also helping to form clouds by large scale gravitational instabilities
(Paper I). The part of it that is in the halo would have no disk or
pressure consequences.

Cosmological dark matter in spiral galaxies would, in principle, have the
same effect, but because the disk is a much smaller fraction of the volume
in the inner regions of spiral galaxies, and the velocity dispersion of
the halo dark matter is a much larger factor times the disk dispersion
(not an independent statement), the impact of this dark matter on the
disk dynamics should be relatively small.

\section{A Problem with Disk Dark Matter}
\label{sect:prob}

Although disk dark matter could, in principle, explain the HII region
pressure anomaly, there is no direct evidence for substantial amounts
of dark matter in galaxy disks, and there is also direct evidence for a
lack of it in the Milky Way, considering the scale heights in the Solar
neighborhood and outer disk (cf. Sect. \ref{sect:ddm}). What can we learn
about disk dark matter in dwarf galaxies from the scale heights there?

\begin{figure}
\vspace{7.0in}
\includegraphics{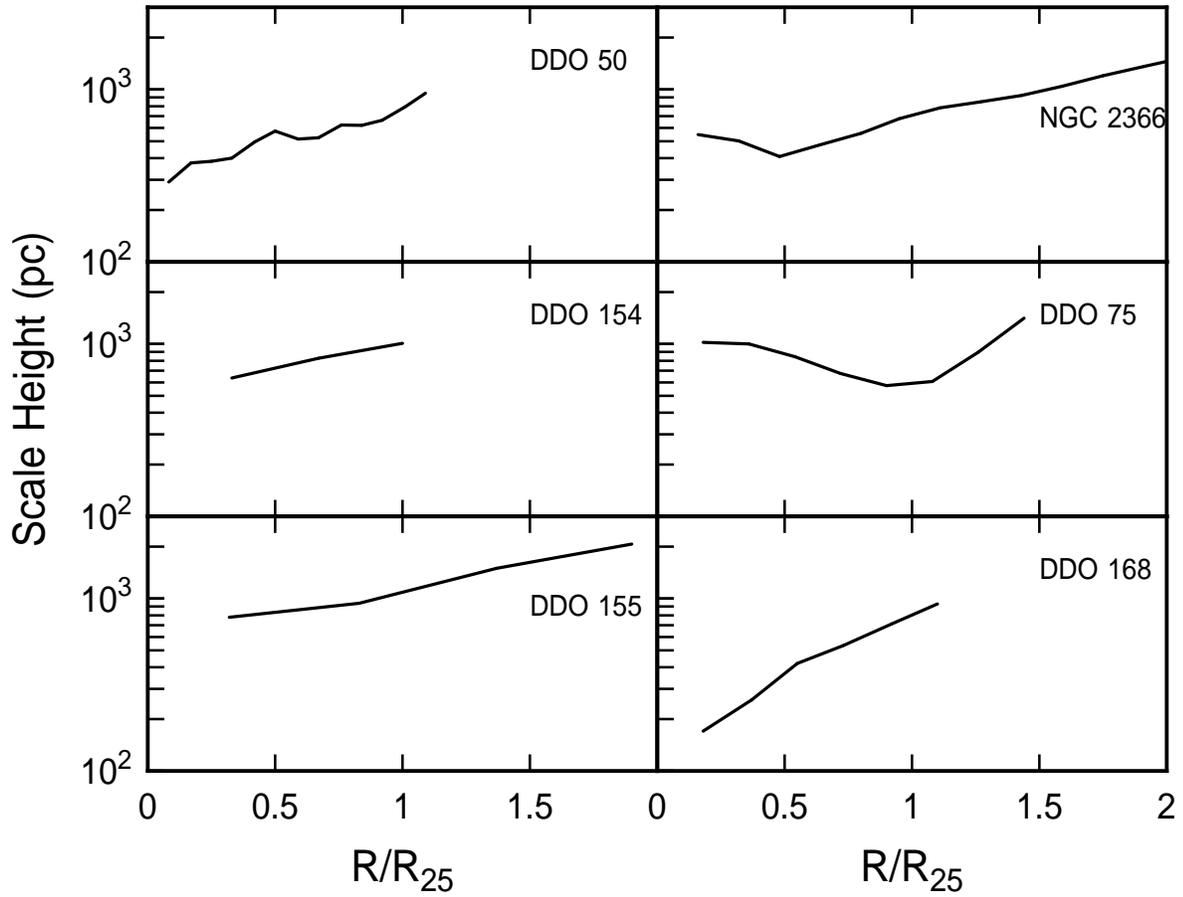}
\caption{Scale height versus radius for the dwarf galaxies, calculated
using a gaseous velocity dispersion of 10 km s$^{-1}$ and the 
observed gaseous and stellar surface densities to obtain the disk
self-gravity.  The theoretical scale heights are comparable to the
observed scale heights, suggesting there is not much 
dark matter in the disks. }
\label{fig:sh}
\end{figure}

Figure \ref{fig:sh} shows the theoretical gaseous scale heights for
our galaxies based on the equation from equilibrium: \begin{equation}
H={{c_{g}^2}\over{\pi G\left[\sigma_{g}+\left(c_g/c_s\right)
\sigma_s\right]}}. \end{equation} The denominator should be the total
column density inside the gas layer, including dark matter if there is
any, but here it is written with only the known gaseous and stellar
disk matter to see if we get a reasonable result in this limit. The
numerator is the total effective gaseous velocity dispersion, including
cosmic rays and magnetic fields, as discussed above. In the figure we use
$c_g=10$ km s$^{-1}$ (van der Kruit, \& Shostak 1982, 1984; Shostak \&
van der Kruit 1984; Dickey, Hanson \& Helou 1990; Boulanger \& Viallefond
1992). If $c_g\sim7$ km s$^{-1}$ without any contribution from magnetic
effects (i.e., as measured directly by the HI velocity dispersion in
some galaxies; van der Kruit \& Shostak 1984; Walter \& Brinks 1999),
the scale heights would be about half the values in the figure.

The figure indicates that the theoretical scale heights range between
$H\sim500-1000$ pc for $c_g\sim10$ km s$^{-1}$ and half of that for
$c_g\sim7$ km s$^{-1}$. These values are comparable to the estimates for
dwarf Irregulars based on bubble size ($H\sim625$ pc for DDO 50 in Puche
et al. 1992; $\sim350$ pc for IC 2574 in Walter \& Brinks 1999), large
disk aspect ratios (Staveley-Smith et al. 1992; Carignan \& Purton 1998),
and direct thickness measurements ($\sim460$ pc for NGC 5023 in Bottema,
Shostak, \& van der Kruit 1986). This implies there is already enough mass
in the disks of these galaxies to account for the gaseous scale heights,
without the addition of any dark matter. It is not possible to increase
the disk dark matter by a factor of $\sim10$ without making the scale
heights too small by the inverse proportion.

Other evidence for negligible disk dark matter was given by Swaters
(1999), who measured the stellar velocity dispersion in the face-on dwarf
galaxy UGC 4325.  He found the dispersion to be too small to allow the
dark matter inferred from the rotation curve to be present in the disk.

The only way substantial disk dark matter can be present and still have
the inferred large scale heights is if the effective velocity dispersion
is large, $20-30$ km s$^{-1}$. This would also increase the ISM pressure
to the value required by HII region near-equilibrium.  However, there
is no evidence for such fast motions in the HI observations, and it is
unlikely that a dwarf galaxy, which has very little shear, can generate a
strong enough magnetic field to give an effective dispersion that large
when the actual rms speed of the gas is only $\sim7-10$ km s$^{-1}$.
Consideration of only a cold ISM component, such as molecular clouds with
a dispersion of $\sim3$ km s$^{-1}$, makes the scale height problem worse.
We conclude that there must be some explanation for the systematically
high HII region pressures that does not rely on either disk dark matter
or high gas velocity dispersions.

\section{Locally high disk pressure and self-gravity 
from giant cloud complexes}
\label{sect:local}

The pressure anomaly found here may be related to the critical column
density anomaly found in Paper I and by van Zee, et al. (1997). In both
cases, the self-gravities of the disks near regions of star formation
are not large enough if the average column densities of the gas and
stars are used to determine them; the higher values given by the local
column density peaks should be used instead. An increase in $\sigma_g$
by a factor of $\sim3-4$ on the scale of the average interstellar Jeans
length would account for both anomalies.

The HI distributions in dwarf Irregulars are indeed clumpy on large
scales. This differs from the morphology of giant spiral galaxies,
where the HI is mostly in the form of spiral arms and in the giant
cloud complexes of these arms. In dwarfs, there may be two or three HI
concentrations in the whole optical disk, and these are the main regions
where stars form.

The peak HI column density in a typical concentration of a dwarf
Irregular is larger than the peak HI column density in a spiral galaxy
(Skillman 1985; Lo, Sargent \& Young 1993; Cayatte et al. 1994;
Broeils \& van Woerden 1994).  HI surveys of dwarfs give peak
column densities in the range of 1 to 3 $\times10^{21}$ cm$^{-2}$ on
scales of several hundred parsecs. In DDO 75 (Sextans A), the peak is
$N(HI)=3\times10^{21}$ cm$^{-2}$ (Skillman, et al. 1988). In DDO 154,
it is $1.6\times10^{21}$ cm$^{-2}$ (Carignan \& Purton 1998), and in
DDO 155, it is $2\times10^{21}$ cm$^{-2}$ (Lo et al., 1993). These
column densities would correspond to several magnitudes of extinction
for Milky Way gas, sufficient to allow the formation of molecules with
self-shielding and dust-shielding. In dwarfs, however, the gas is still
HI (presumably because of the low metal abundance), although there may
be additional gas that is H$_2$ also.

The peak HI mass column density in dwarf Irregulars is also larger
than the stellar mass column density (Lo et al. 1993). This
means that the pressure is easily estimated from the equation
$P=\left(\pi/2\right)G\sigma_g^2$ where $\sigma_g=2.2\times10^{-24}N(HI)$
gm cm$^{-2}$, considering He and heavy elements. This is a lower limit
to the mid-disk pressure because $N(HI)$ is a lower limit to the column
density if there is also H$_2$ in the clouds. For the observed range of
peak column densities, $N(HI)=1-3\times10^{21}$ cm$^{-2}$, the pressure
is in the range $3.7\times10^3-3.3\times10^4$ K cm$^{-3}$.  The upper
range is about what is observed for the HII regions in Figure \ref{fig:p}.

This result implies that a likely explanation for the relatively high
HII region pressures shown in Figure \ref{fig:p} is a relatively high
{\it local} column density of gas on the 100--300 pc scale of the
star formation regions. The {\it average} HI column density in a dwarf
Irregular is comparable to that in a giant spiral, but the large-scale
peaks are much higher than the average in the dwarfs, and the valleys
between the peaks are much lower than average.

Most of the HII regions in our dwarf galaxy sample are indeed where the
HI column density exceeds $10^{21}$ cm$^{-2}$.  For NGC 2366, our HI data
(Hunter, Elmegreen, \& van Woerden 2000) has a resolution of 530 pc,
which is too large to make a detailed comparison, but the supergiant
HII region is clearly resolved on this scale, and it sits within an
HI contour of $\sim4\times10^{21}$ cm$^{-2}$. The other giant HII
region to the west is to the side of an HI peak, but within a contour
of $3\times10^{21}$ cm$^{-2}$.  The collection of HII regions to the
northeast sit around HI peaks and within a contour of $3\times10^{21}$
cm$^{-2}$.  There is one HII region there that does not sit on a peak
but is still within the $3\times10^{21}$ cm$^{-2}$ contour.  A lone HII
region to the extreme northeast is within a contour of $3\times10^{21}$
cm$^{-2}$.  Off to the southwest there are two faint small HII regions
that sit in a contour of $1\times10^{21}$ cm$^{-2}$.  Thus, all but a
few small HII regions are within an HI contour of about $3\times10^{21}$
cm$^{-2}$.  There the pressure from the self-gravity of the {\it local}
gas alone exceeds $3\times10^4$ K cm$^{-3}$, and can easily account for
the confinement of the HII regions.

Similarly in DDO 50, the HII regions are in the HI clumps rather than
in any of the HI holes, according to Figure 21 in Puche et al. (1992).
Their HI beam size was 70 pc, but there are no contours for $N(HI)$
in Puche et al., so we cannot determine the self-gravitating pressures.

For DDO 75, Figure 5 in Hodge, Kennicutt, \& Strobel (1994) shows
the HII regions sketched on the HI contours of Skillman et al. (1988),
where the beam size was 340 pc.  A big HII complex to the east is offset
from the center but still closely associated with an HI cloud, within
a contour of about $1.5\times10^{21}$ cm$^{-2}$.  The same is true for
an H$\alpha$ shell to the west, which is in another distinct HI cloud.
Hodge et al. identify some HII regions between these two HI clouds,
but our H$\alpha$ image suggests they are filaments associated with the
big HII complex. There are also smaller HII regions to the northeast of
the big HII complex, but they also fall within the same big HI cloud.

In DDO 154, Carignan \& Beaulieu (1989) mapped HI with a beam size of 870
pc. Their HI contour map (Fig. 8) is rather featureless, but it appears as
if most of the HII regions lie within a contour of about $1\times10^{21}$
cm$^{-2}$.  Carignan, Beaulieu, \& Freeman (1990) also studied DDO 155
in HI with a beam size of 107 pc. We compared our H$\alpha$ image to
their HI map, and it appears that the HII regions are associated with
HI peaks having $1\times10^{21}$ cm$^{-2}$.  DDO 168 was mapped in HI
by Broeils (1992) with a beam size of 320 pc. The HI contours on the
optical image suggest that the HII regions are all within a contour of
$2\times10^{21}$ cm$^{-2}$.

These HI column densities should be considered lower limits because of
the additional molecular mass that is likely to be present in the HI cloud
cores, because the HI can be optically thick in a cool center, and because
the peaks in the HI are washed out by the relatively low resolution
of most of the HI maps. The implication is that {\it the pressures in
the star-forming regions of dwarf Irregular galaxies are determined
primarily by the self-gravities of the local HI concentrations}. These
pressures are comparable to those in the star-forming regions of giant
spiral galaxies even though the background stellar disks have much lower
column densities in the Irregulars, and even though the {\it average}
pressures are much lower in the Irregulars.

The dominance of local gravity in regions of star formation would make
these regions somewhat autonomous in the galactic environment.  They would
be like mini-galaxies inside the overall dwarf galaxy, possibly producing
a patchy metallicity distribution if the wind and supernova ejecta get
partially trapped.  They could even remain coherent on a cosmological time
scale, migrating around the disk, perhaps even into the nuclear region,
where by continued star formation, they eventually make a bright nuclear
cluster as in a nucleated dwarf elliptical (Sandage \& Hoffman 1991;
Phillips, et al. 1996).

The proposed increase in the local HI column density over the
average used for equation (\ref{eq:p}) implies a lower scale height
in regions of star formation than elsewhere. This would seem to
contradict the discussion in Section \ref{sect:prob} about the
need for a scale height in the range of $300-500$ pc, but this is
not the case. The density at a particular height above the midplane
does not decrease much if the scale height decreases by a factor of
$\sim2$ following an increase in the local disk column density. This
is because the mass density at a particular height $z$ is given by
$\left(\sigma/2H\right)e^{-0.5\left(z/H\right)^2}$ for mass column density
$\sigma$ and scale height $H=c^2/\left(\pi G\sigma\right)$. If $\sigma$
increases, then $H$ decreases, making the exponential factor smaller,
but the term in front of the exponential increases to compensate. For
example, suppose the density at a height of 300 pc is required to be
large in order to contain the observed giant shells and to account
for the large inferred galaxy aspect ratios, as discussed in Section
\ref{sect:prob}. If the scale height is assumed to be 300 pc in order to
account for this high density using the average column density, then we
can calculate what the density at that same height will be if the column
density increases by various factors. The relative density at a fixed
height varies with column density as $\left(\sigma/\sigma_0\right)^2
e^{0.5\left(1-\sigma^2/\sigma_0^2\right)}$, where the normalization
gives the density relative to the value at one scale height before the
variation in $\sigma$. For $\sigma/\sigma_0=$ 1, 1.5, 2, 2.5, and 3, the
density at the fixed height equals 1, 1.2, 0.89, 0.45, and 0.16. Thus,
the gas column density all around a region of star formation can be
higher than the galactic average at that radius by a factor of $\sim3$
without seriously affecting the ability of the gas to contain giant
shells, and without greatly affecting the galaxy aspect ratio at a {\it
fixed} density.

\section{Overpressure and Continued HII Region Expansion: The Origin of
Giant Shells?}
\label{sect:shell}

The similarity between the detection limit for HII regions and the lower
limit of the densities plotted in Figure \ref{fig:p} suggests that there
are probably more HII regions that are too faint to see. These would
have lower pressures that could conceivably come all the way down to
the average disk pressure.

The radius $r$ of a Str\"omgren sphere expanding
into a uniform density increases with time $t$ as
$r=r_0\left(1+7c_{HII}t/\left[4r_0\right]\right)^{4/7}$ for sound
speed $c_{HII}$ and initial radius $r_0$.  For very late times,
$r\propto t^{4/7}$, and since $r\propto n^{-2/3}$ to contain all of the
recombination, $t\propto n^{-7/6}$.  This implies that the time required
for the density to drop by another factor of $\sim10$ to reach pressure
equilibrium is $\sim15$ times longer than the average age of the oldest
HII regions in the present study.  Consequently, in a steady state, we
could be missing 93\% of the HII regions below the detection limit if
they stay ionized that long. Moreover the volume of a Str\"omgren sphere
increases as $n^{-2}$, so the summed volumes of all the HII regions
could be larger than the observed volume by a factor of $\sim100$.
This summed volume is enough to fill the entire disk and spill out into
the halo.  It follows that if the ISM in a dwarf galaxy has a pressure
as low as the average disk pressure calculated in Section \ref{sect:p},
then common HII regions should expand and merge to fill the entire disk.

The continued expansion of HII regions as a result of their large total
pressures from ionization, supernovae, and stellar winds would lead to the
slow creation of giant shells that have such a large volume filling factor
that they could cover an entire dwarf galaxy. Shear will not distort
these shells much because the rate of shear is relatively low.  The disk
will therefore be in a permanent state of severe disruption, possibly
resembling the shell-filled galaxy Ho II (= DDO50; Puche et al. 1992).
A significant amount of material should also flow into the halo, since
the disk self-gravity is low if the average disk column density is low.

The ages of the giant shells in HoII could be large, perhaps $\sim10$
times larger than the ages of the HII regions now seen in the disks if
they continue to expand in a snowplow phase after the OB stars leave the
main sequence.  Such large ages, approaching $\sim50$ My, imply that
the central OB associations would have dispersed because of random
stellar motions, and they would have lost their most massive stars
because of stellar evolution.  This appears to have been the case for
the giant shell LMC4 in the Large Magellanic Cloud, which may have begun
its expansion $\sim30$ My ago at the site of what is now a dispersed
cluster of A stars (Efremov \& Elmegreen 1998).  The lack of detection
of OB associations or other clusters inside the shells of HoII (Rhode,
et al. 1999) does not imply these shells had peculiar pressure sources.
Common stellar pressures could have made them, as observed in some cases
(Steward, et al. 2000).  Such shells are expected if the average ISM
pressure is $\sim10$ times lower than the visible HII region pressures
in these galaxies.

\section{A Comparison Between Dwarf Irregulars and Giant Spirals}
\label{sect:spiral}

Section \ref{sect:local} suggested that dwarf galaxies have large
pressure fluctuations in their disks resulting from highly irregular
gas distributions, and that star formation systematically occurs at the
positions of the giant cloud complexes, where the gas column density is
locally high. Such clouds can confine the HII regions and give them a
higher pressure than the average ISM at that radius, and they should also
help to confine any giant shells that might be produced there, making
them appear round even at a radius of several hundred parsecs. Such
large-scale irregularities in the gas distribution make sense for dwarf
galaxies, which are also irregular in their star formation, but why isn't
there a similar irregularity in giant spirals?  Section \ref{sect:shell}
also suggested that without such systematic local confinement of the HII
regions, there should arise a pervasive network of giant shells, filling
the entire galaxy.  What prevents this from happening in spiral galaxies?

There are three primary differences between dwarf Irregulars and giant
spiral disks with regard to their ISM kinematics: the spirals have
larger average surface brightnesses, they often have strong density
waves, and they always have shear. These three differences contribute
toward our perception that the pressures, threshold column densities,
and gas morphologies are anomalous in dwarfs. Without shear or spiral
waves, the giant clouds that are formed by turbulence compression
and gravitational instabilities can remain for a long time as clumpy
objects with locally high pressures and locally strong self-gravity.
With the shear in a larger galaxy, these clouds are rapidly stretched
and distorted.  Pressure-driven shells can also become very large
without shear and with the low background pressures implied by the
low surface brightnesses of dwarfs.  If there is no density wave, or a
wave is very weak as in a flocculent galaxy, then the sheared clouds and
their associated star formation take the form of short irregular spirals
separated by low-density interspiral regions. In flocculent spiral arms,
the peak gaseous column density and the pressure can be high, and the
self-gravity can be strong, allowing stars and normal HII regions to
form. Between the arms the pressure is low.  This is similar to the
situation in dwarfs except that the larger sizes of the spiral galaxies
lead to more clouds and the shear makes these clouds spiral-like, but
the pressure variations are the same.

When there are density waves in a large spiral galaxy, the wave
modulates the overall gas dynamics and again prevents the gas from
clumping into long-lived complexes. However, the way it does this is
different than for flocculent galaxies. Inside a density wave arm,
the shear is locally low, and the clouds and shells that form can avoid
distortion. The clouds remain like clumps in the usual beads-on-a-string
pattern (Elmegreen 1994) and the shells can become large, like several
in the southern arm of M83. Spiral arm clumps are similar in size,
mass, and gravitational self-binding to the gas concentrations in dwarf
Irregulars (aside from overall scaling differences for the Jeans length --
Elmegreen et al. 1996), but they are short-lived in the spirals because
the flow-through time in the arm is short. The interarm region is a
hostile environment for large clouds and shells: it tears them apart
with heightened tidal forces and greater-than-average shear.

\section{Warm and Cool Phases of Interstellar Gas}

Another implication of a large-scale clumpy structure of HI in dwarf
galaxies concerns the phase of the gas, i.e., whether cool clouds
are allowed to exist in thermal equilibrium as a precursor to star
formation. When the pressure is low, as in the outer parts of galaxies
and in low surface brightness galaxies, the neutral interstellar gas
is mostly in the warm phase; there is no cool phase except in regions
with locally high pressures (Elmegreen \& Parravano 1994). In fact,
the outer regions of spiral galaxies are observed to be mostly in this
pure-warm phase (Braun 1997). This means that star formation is not
likely to begin in the extremely low pressure environment of a dwarf
galaxy until a giant HI concentration forms, increasing the pressure
up to a value that supports cool HI. This difficulty in forming stars
explains the problem raised by Lo, Sargent, \& Young (1993), namely how
dwarf galaxies can remain so long at a low surface brightness without
converting their abundant HI into stars. The solution is that the gas is
usually in the warm phase and has to clump up gravitationally first in
order to permit a cool phase of the neutral medium to form. High pressure,
strong self-gravity, and star formation then follow. This is presumably
why the star formation and HII regions in dwarf galaxies look remarkably
normal even though the average galaxy is different from the Milky Way.

The existence of much of the HI gas in a warm state except where it is
associated with star formation is consistent with the observations of
DDO 69 (Leo A) by Young \& Lo (1996). They found that there were two
components of the HI: a broad-lined ($\sigma=$9 km s$^{-1}$) component
that was observed everywhere and a narrower component ($\sigma=$3.5 km
s$^{-1}$) near HII regions. They interpreted these two components as a
warm and cold phase of the neutral gas.  These two components are also
observed in Sag DIG (Young \& Lo 1997).  The prevalence of the warm phase
in dwarf galaxies also explains the anomalously high HI spin temperatures
of HI (eg., Lo, Sargent, \& Young 1993). Most of the gas on the periphery
of a giant HI concentration and elsewhere in the disk is in the warm
phase. Only the cores of the condensations have a high enough pressure
to be cool. These cores should be highly molecular as well.

\section{Conclusions}

Near-equilibrium between the average pressures of old HII regions and
the back-pressure of the local interstellar medium requires that dwarf
galaxies have a larger disk column density in the vicinity of the star
formation than the average at that radius observed in the form of HI or
starlight emission. The mass enhancement is a factor of $\sim3$ if the
unseen material is gas-like and able to contribute to the ISM pressure
directly, and a factor of $\sim10$ if it is star-like or in the form of
non-interacting cosmological particles, able to contribute only to the
disk self-gravity.

A factor of $\sim10$ enhancement in disk dark matter is close to the
factor required to bind dwarf galaxies overall, as inferred from the
rotation curves or virial velocities. This implies that a large fraction
of the dark matter in a dwarf galaxy could be in the disk, contributing to
the disk self-gravity in the perpendicular direction, and therefore to the
overall gas pressure (through its effect on the scale height of the gas).

A direct measurement of such a contribution to the scale heights of
these galaxies gave a negative result, however. The dark matter that
contributes to the rotation cannot be in the disk or the scale height and
disk aspect ratio would be much smaller than the observations permit. The
dark matter is therefore in the halo, and the high pressure that may be
confining the HII regions has to have another source.

We suggested that this source is the self-gravitational binding pressure
from the clouds that formed the stars in the HII regions. Most of the
gas in the optical disks of dwarf Irregular galaxies is concentrated in
a small number of relatively large HI clouds. The cores of these clouds
have large enough pressures to explain the anomaly with the HII regions.
They can also account for other peculiarities about star formation in
dwarfs, including the low average surface density threshold, and the
suspected inability of the gas to make cool neutral clouds in a low
pressure environment.

HII regions that are not confined by the large pressure from a
surrounding cloud can expand into the low ambient pressure for a long
time. Considering how far they have to go to reach pressure equilibrium
in dwarf galaxies, there could be 10 times as many HII regions in dwarf
galaxies than what we observe here, and their total volume filling factor
could approach 100\%.  If this is the case, then the numerous giant shells
observed in Ho II (DDO50) would be old expanded HII regions and wind-swept
bubbles from star formation that took place as long as $\sim50$ My ago.
The associations would be difficult to see now.

Consideration of the differences between dwarfs and giant spiral
galaxies clarifies the importance of pressure variations somewhat
further. In flocculent spirals, giant clouds that may be analogous to
the HI concentrations in dwarfs get sheared into spiral-like shapes,
after which they are not readily recognized as having the same basic
properties and formation mechanisms. In spirals with density waves,
the high pressure HI is primarily in the arms. A lack of shear at the
arm crests gives the HI concentrations a similar globular form to the
giant HI clouds in dwarfs, but not necessarily a similar life time.
Giant clouds are relatively short lived in spiral arms because of the
changing pressure from the wave motion. They could in principle live
much longer in dwarf galaxies.

A potentially important application of these results concerns the
dispersal of the heavy elements that are made in the star-forming regions
of dwarf Irregular galaxies. If the gravitational potential wells of
these regions are as deep as those inferred here from the local HI
column densities and the HII region pressures, then this dispersal may
not be as straightforward as was once thought. Numerical modeling is
required to determine what luminosities are needed for supernova- and
wind-swept bubbles to break out of the high-pressure gas concentrations
around each star-forming region, and get into the galaxy halo or the
intergalactic medium.

\acknowledgments

Support for this work was provided to D.A.H. by grant AST-9802193 and
to B.G.E. by grant AST-9870112 from the National Science Foundation.
Helpful comments by the referee are gratefully appreciated.

\clearpage
 
\begin{deluxetable}{lrrrrr}
\scriptsize
\tablecaption{Galaxy Parameters and References \label{tabref}}
\tablewidth{0pt}
\tablehead{
\colhead{Galaxy}
& \colhead{pa\tablenotemark{a}} & \colhead{i\tablenotemark{a}}
& \colhead{$c$\tablenotemark{b}}
& \colhead{HI Reference} & \colhead{$\mu$ Reference} \nl
\colhead{}
& \colhead{(\arcdeg)} & \colhead{(\arcdeg)}
& \colhead{(km/s)}
& \colhead{} & \colhead{}
}
\startdata
DDO 50
&  30 & 40 & 6.8\xx &
Puche \et\ 1992 & Hunter, Elmegreen, \& Baker 1998 \nl
DDO 75
& 50 & 36 & 9.0\xx &
Skillman \et\ 1988 & Hunter \& Plummer 1996 \nl
DDO 154
& 38 & 57 & (9)\xx &
Carignan \& Beaulieu 1989 & Carignan \& Beaulieu 1989 \nl
DDO 155
&  47 & 47 & 9.5\xx &
Carignan \et\ 1990 & Carignan \et\ 1990 \nl
DDO 168
& 328 & 76 & (9)\xx &
Broeils 1992 & Broeils 1992 \nl
NGC 2366
& 32 & 73 & 15\xx &
Hunter, Elmegreen, \& van Woerden 2000 & Hunter, Elmegreen, \& van
Woerden 2000 \nl
\enddata
\tablenotetext{a}{Position angles and inclinations are taken from the
HI references. 
The exception is DDO 50 where the HI position angle did not make sense
for the optical image; instead we used a position angle determined
from the broad-band optical image.}
\tablenotetext{b}{Velocity dispersion of the gas. Values in parenthesis
are assumed; values for DDO 155 and DDO 50 are galaxy-wide averages.}
\end{deluxetable}

\begin{deluxetable}{lccc}
\tablecaption{Ratio of Total Pressure to rms Thermal Pressure in a Turbulent HII Region}
\tablehead{
\colhead{}
&\colhead{M=3.16}&\colhead{5}&\colhead{10}\nl
\colhead{f}&\colhead{}&\colhead{}&\colhead{}
}
\startdata
1&11&26&101\nl
0.316&7.4&16&58\nl
0.1&6.3&11&35\nl
0.0316&7.0&10&23\nl
0.01&8.5&12&20\nl
\enddata
\end{deluxetable}

\begin{deluxetable}{lccc}
\tablecaption{Average Thermal Pressures in HII Regions and Disk beyond $0.5R_{25}$}
\tablehead{
\colhead{Galaxy}
&\colhead{$<$P(HII)$>$}
&\colhead{$<$P(disk)$>$}
&\colhead{${{<{\rm P(HII)}>}\over{<{\rm P(disk)}>}}$}\nl
\colhead{}
&\colhead{($\times10^3$ K cm$^{-3}$)}
&\colhead{($\times10^3$ K cm$^{-3}$)}&\colhead{}
}
\startdata
DDO50&30.2&4.3&7.1\nl
DDO75&35.2&3.2&11.0\nl
DDO154&10.8&2.1&5.2\nl
DDO155&41.9&0.8&55\nl
DDO168&22.2&4.9&4.5\nl
NGC 2366&29.6&2.0&14.8\nl
\enddata
\end{deluxetable}

\clearpage
 
\clearpage

\end{document}